\documentclass[aps,prd,10pt,reprint,nofootinbib,floatfix]{revtex4-1}

\usepackage{graphicx}
\usepackage{dcolumn}
\usepackage{amssymb}
\usepackage{amsmath}
\usepackage{amsfonts}
\usepackage{amsbsy}
\usepackage{color}
\usepackage{rotating}
\usepackage{lipsum}
\usepackage[english]{babel}
\usepackage{soul}
\usepackage{layouts}
\usepackage{multirow}
\usepackage{comment}
\usepackage{overpic}
\usepackage{silence}
\usepackage[english]{babel}
\usepackage[utf8]{inputenc}
\usepackage[dvipsnames]{xcolor}

\usepackage{soul}
\usepackage{cancel}
\usepackage{verbatim}
\usepackage{lipsum}

\usepackage[colorlinks]{hyperref}
\AtBeginDocument{\hypersetup{
        linkcolor=Blue,
        allcolors=Blue
}}
\usepackage{orcidlink}

\newcommand*{\rom}[1]{\expandafter\@slowromancap\romannumeral #1@}
\makeatother


\begin{document}

\title{Generalizing the CPL Parametrization through Dark Sector Interaction}

\author{Mikel Artola\orcidlink{0009-0007-9068-1995}} 
\email[Contact author: ]{mikel.artola@ehu.eus}
\author{Ruth Lazkoz\orcidlink{0000-0001-5536-3130}}
\email[Contact author: ]{ruth.lazkoz@ehu.eus}
\affiliation{Department of Physics, \href{https://ror.org/000xsnr85}{University of the Basque Country EHU}, 48940 Leioa, Spain}
\affiliation{EHU Quantum Center, \href{https://ror.org/000xsnr85}{University of the Basque Country EHU}, Leioa, 48940 Leioa, Spain}
\author{Vincenzo Salzano\orcidlink{0000-0002-4905-1541}}
\email[Contact author: ]{vincenzo.salzano@usz.edu.pl}
\affiliation{Institute of Physics, \href{https://ror.org/05vmz5070}{University of Szczecin}, Wielkopolska 15, 70-451 Szczecin, Poland}

\begin{abstract}
We investigate a hierarchy of interacting dark energy (IDE) models featuring a non-gravitational coupling between dark matter and dark energy. Specifically, we examine scenarios where the background interaction kernel, $Q = 3H(\delta + \eta a)\rho_\mathrm{de}$, allows for both constant and dynamical coupling parameters. Adopting the Chevallier-Polarski-Linder parametrization for the dark energy equation of state, $w_\mathrm{de} = w_0 + w_a(1-a)$, we derive closed analytical expressions for the energy densities of dark matter and dark energy. Afterwards, we obtain observational constraints using joint combinations of DESI DR2 baryon acoustic oscillations, Pantheon$+$ Type Ia supernovae, and Planck$+$ACT compressed cosmic microwave background likelihoods. For constant coupling models, we find parametric deviations from $\Lambda$ ranging from $2.7\sigma$ to $2.9\sigma$; however, for interactions with dynamical couplings, these significances are reduced to $1.3\sigma$--$1.5\sigma$. Ultimately, our Bayesian model comparison reveals that no investigated IDE scenario is statistically preferred over the concordance $\Lambda$CDM model. These results highlight the necessity of reporting Bayesian evidence alongside conventional frequentist maximum-likelihood analyses to ensure robust cosmological claims concerning dark energy evolution and interaction.


\end{abstract}

\maketitle

\section{Introduction}

Gravity, though the weakest of the fundamental interactions, governs the formation, evolution, and ultimate fate of the largest structures of the Universe. General Relativity~\cite{Einstein:1915ca, Einstein:1915by} has established itself as the most accurate and comprehensive description of gravitational phenomena, and has withstood rigorous testing across a vast range of scales---from local Solar System experiments to observations probing cosmological distances~\cite{Will:2014kxa, LIGOScientific:2021sio}. Yet, on these largest scales, explaining the observed expansion history and the growth of structure within General Relativity requires the introduction of two dominant but enigmatic components: dark matter (DM) and dark energy (DE). In the standard $\Lambda$CDM framework, DM is characterized as cold (a non-relativistic and collisionless component that acts as the main gravitational source of structure formation), while DE is represented by a cosmological constant $\Lambda$ that drives the current accelerated expansion of the Universe. The latter was first revealed by Type Ia supernovae (SN)~\cite{SupernovaSearchTeam:1998fmf, SupernovaCosmologyProject:1998vns}, and ever since its existence has been reliably backed up by precision measurements of baryon acoustic oscillations (BAO)~\cite{SDSS:2005xqv, 2dFGRS:2005yhx} and the cosmic microwave background (CMB)~\cite{Boomerang:2000wrj, WMAP:2003gmp}.

Despite its success, the concordance $\Lambda$CDM model faces a mounting array of observational and theoretical challenges. Leading the observational hurdles are cosmological tensions brought by recent high-precision data. The most prominent is the Hubble tension~\cite{Planck:2018vyg, Riess:2021jrx, SPT-3G:2025bzu, H0DN:2025lyy}, often characterized as a severe discrepancy between the value of the Hubble constant $H_0$ inferred from early-Universe probes and that obtained from local measurements~\cite{Pantos:2026cxv}. Similarly, the $S_8$ tension reveals an ``early--late'' inconsistency regarding the amplitude of matter clustering~\cite{Erben:2013aaa, Planck:2013pxb, Kilo-DegreeSurvey:2023gfr, Wright:2025xka}. Furthermore, recent results from the Dark Energy Spectroscopic Instrument (DESI)~\cite{DESI:2024mwx, DESI:2025zgx} have placed additional strain on the cosmological constant paradigm, with data indicating a mild preference for dynamical DE and a recent crossing of the phantom divide in its equation-of-state parameter. Even so, recent Bayesian analyses~\cite{Ong:2025utx, Ong:2026tta} continue to support the non-dynamical nature of DE, so the debate continues. For a comprehensive review of these and other emerging discrepancies, we refer the reader to~\cite{CosmoVerseNetwork:2025alb} and the references therein.

Beyond these empirical obstacles, the concordance model is not exempt from serious theoretical issues. The observed value of $\Lambda$ differs from quantum field theory predictions by roughly 120 orders of magnitude, necessitating extreme fine-tuning and giving rise to the long-standing cosmological constant problem~\cite{Weinberg:1988cp, Martin:2012bt}. A related problem is the so-called coincidence problem~\cite{Zlatev:1998tr}: the mystery of why the energy densities of DM and DE are of the same order at the current cosmic time. Given their vastly different dilution rates---$\rho_\mathrm{dm} \propto a^{-3}$ vs.~$\rho_\Lambda = \mathrm{constant}$---their densities should differ by many orders of magnitude throughout most of cosmic history, yet they happen to be of the same order today.

The aforementioned concerns motivate the search for alternative cosmological models capable of addressing these substantial challenges. Given the unknown nature of both the DM and DE sectors, interacting dark energy (IDE) models have gained significant attention by postulating non-gravitational interactions between these components. Although the coincidence problem provided the initial motivation for these studies~\cite{Amendola:1999er, Zimdahl:2001ar, Comelli:2003cv, Chimento:2003iea, Farrar:2003uw, Wang:2004cp, Olivares:2006jr, Sadjadi:2006qp, Quartin:2008px, delCampo:2008jx, Caldera-Cabral:2008yyo, He:2010im, delCampo:2015vha, vonMarttens:2018iav}, IDE scenarios have since emerged as compelling candidates to alleviate cosmological tensions (see \cite{vanderWesthuizen:2025vcb, vanderWesthuizen:2025mnw, vanderWesthuizen:2025rip} and references therein for an exhaustive list, as well as~\cite{Wang:2025znm, Wang:2026wrk}). Furthermore, these models provide a natural physical mechanism to explain the potentially dynamic nature of the dark sector~\cite{Benisty:2024lmj, Li:2024qso, Guedezounme:2025wav, Li:2026xaz}.

In this study, we investigate a class of phenomenological IDE models at the background level. Specifically, we consider an interaction kernel proportional to the energy density of DE that incorporates a dynamical coupling evolving with cosmic expansion. To remain agnostic about the underlying nature of DE, we further assume that its equation-of-state parameter $w_\mathrm{de}$ is also dynamical. We exclude any interactions extending beyond the dark sector, motivated by the tight constraints imposed by local gravity measurements and the polarization of radiation from distant sources~\cite{Carroll:1998zi, ParticleDataGroup:2002ivw, Peebles:2002gy, Wang:2016lxa}.

The structure of this work is organized as follows. In Section~\ref{sec:IDE}, we derive the exact analytical solutions for the energy densities of DM and DE under a linear coupling with the scale factor and a dynamical CPL equation of state. We also establish the theoretically motivated constraints on the dark sector parameter space to ensure dark energy subdominance at early times and stability within the perturbative sector. Section~\ref{sec:methodology} presents the observational probes and the Bayesian methodology employed for parameter estimation. Subsequently, Section~\ref{sec:results} is devoted to the presentation and discussion of our findings. We conclude with our final remarks in Section~\ref{sec:conclusions}.

\section{\label{sec:IDE} Interacting Dark Energy Models}

\subsection{Theoretical framework and model}

Throughout this work, we operate within the framework of GR, assuming a spatially flat Friedmann-Lemaître-Robertson-Walker (FLRW) background~\cite{Planck:2019nip}. Adopting natural units where $c \!=\! 1$, the line element is given by
\begin{equation}
    \mathrm{d}s^2 =
    -\mathrm{d}t^2 + a(t)^2 \delta_{ij} \mathrm{d} x^i \mathrm{d}x^j,
\end{equation}
where $t$ is the cosmic time, $a(t)$ is the scale factor, which we normalize to unity at the present time ($a_0 = 1$), and $x^i = (x,y,z)$ are the comoving Cartesian coordinates. The energy content of the Universe is modeled as a set of perfect fluids---radiation, baryons, pressureless DM, and DE---, each characterized by its energy density $\rho_i$, pressure $p_i$, and four-velocity $u_i^\mu$.

While the standard model assumes that these components evolve independently, we investigate a scenario where the dark sector components (DM and DE) interact non-gravitationally. The total stress-energy tensor is conserved ($\nabla_\mu T_\mathrm{tot}^{\mu\nu} = 0$), but the conservation laws for the individual dark components are modified by a source term $Q^\nu$:
\begin{equation}
    \nabla_\mu T_\mathrm{dm}^{\mu\nu} = - \nabla_\mu T_\mathrm{de}^{\mu\nu} = Q^\nu.
    \label{eq:continuity_dark_sector}
\end{equation}
In IDE models, the interaction 4-vector $Q^\nu$ is not derived from a fundamental theory, but is typically introduced phenomenologically. Although specific frameworks can be formulated at the Lagrangian level and motivate particular forms of the interactions (see~\cite{Bolotin:2013jpa} for an extensive review), no underlying principle uniquely determines its functional form, which remains essentially arbitrary, reflecting the absence of a fundamental underlying theory.

In the FLRW background, the interaction can be fully characterized by its temporal component in the comoving frame, $Q$, representing the energy transfer between the dark sector components~\cite{Kodama:1984ziu, Malik:2002jb}. The background dynamics are then governed by the Friedmann equation:
\begin{equation}
    3H^2 = 8\pi G \sum_i \rho_i,
\end{equation}
where $H \equiv \dot{a}/a$ is the Hubble function and the overdot stands for derivation with respect to $t$, as well as the system of coupled continuity equations:
\begin{align}
    \dot{\rho}_\mathrm{dm} + 3H\rho_\mathrm{dm} &= Q, \label{eq:continuity_dm} \\
    \dot{\rho}_\mathrm{de} + 3H(1 + w_\mathrm{de})\rho_\mathrm{de} &= -Q, \label{eq:continuity_de}
\end{align}
where $w_i := p_i/\rho_i$ is the equation-of-state parameter of the $i$th species. Radiation and baryons are assumed to be independently conserved, with equations of state $w_\mathrm{r} = 1/3$ and $w_\mathrm{b} = 0$, respectively. The direction of energy flow in the dark sector is determined by the sign of $Q$: $Q > 0$ implies energy transfer from DE to DM, whereas $Q < 0$ implies the reverse.

Phenomenological interaction functions affecting the background dynamics are frequently constructed as $Q = 3H \alpha f(\rho_\mathrm{DM}, \rho_\mathrm{DE})$, where $\alpha$ is the coupling parameter that characterizes the strength of the interaction, and $f(\rho_\mathrm{dm}, \rho_\mathrm{de})$ is any continuous function of the energy densities of the dark sector components. While most studies assume a constant $\alpha$ for simplicity, it has been argued that there is no fundamental symmetry that restricts the coupling to be a constant or zero~\cite{Li:2011ga, Wang:2014xca, Yang:2019uzo, Yang:2020tax}. Motivated by this, in this work we study an interaction kernel $Q$ proportional to $\rho_\mathrm{de}$, where the coupling strength follows a first-order Taylor expansion in the scale factor $a$:
\begin{equation}
    Q = 3H (\delta + \eta a) \rho_\mathrm{de}.
    \label{eq:Q_linear}
\end{equation}
Here, $\delta$ and $\eta$ are dimensionless constants quantifying the interaction strength. To maintain generality regarding the nature of the DE fluid, we combine this interaction kernel with the Chevallier-Polarski-Linder (CPL) parametrization~\cite{Chevallier:2000qy, Linder:2002et} for the equation of state of DE:
\begin{equation}
    w_\mathrm{de}(a) =
    w_0 + w_a(1 - a).
    \label{eq:CPL}
\end{equation}
This configuration constitutes a comprehensive extension of previously explored interacting scenarios. While observational constraints for related models were addressed prior to the paradigm shift marked by the recent DESI results~\cite{Yang:2022csz}, we advance the theoretical framework by deriving closed analytical expressions for the background energy densities. These solutions, detailed in the following subsection, allow for a more efficient exploration of the dark sector parameter space.

\subsection{Analytical solution for energy densities}

By substituting the ansatz~\eqref{eq:Q_linear} into the continuity equations, we derive closed analytical forms for the energy densities. We first define an effective equation of state for dark energy, $w_\mathrm{de}^\mathrm{eff} = w_\mathrm{de} + Q/(3H\rho_\mathrm{de})$, which, given our linear coupling, simplifies to:
\begin{equation}
    w_\mathrm{de}^\mathrm{eff}(a) =
    (w_0 + w_a + \delta) + (\eta - w_a) a.
    \label{eq:w_eff_de}
\end{equation}
In terms of $w_\mathrm{de}^\mathrm{eff}$, the DE continuity equation~\eqref{eq:continuity_dm} adopts the standard uncoupled form:
\begin{equation}
    \dot{\rho}_\mathrm{de} + 3H(1 + w_\mathrm{de}^\mathrm{eff}) \rho_\mathrm{de} = 0.
\end{equation}
Integrating this equation yields
\begin{equation}
    \rho_\mathrm{de}(a) =
    \rho_\mathrm{de,0} a^{-3(1 + w_0 + w_a + \delta)} e^{3(\eta - w_a) (1 - a)},
    \label{eq:rho_de}
\end{equation}
where $\rho_\mathrm{de,0}$ denotes the current value of the energy density of DE. 

Turning to the DM component, Eq.~\eqref{eq:continuity_dm} can be rewritten as a first-order linear differential equation in terms of the scale factor,
\begin{equation}
    a \frac{\mathrm{d} \rho_\mathrm{dm}}{\mathrm{d} a} + 3\rho_\mathrm{dm} =
    3(\delta + \eta a) \rho_\mathrm{de}(a),
\end{equation}
whose solution in terms of a quadrature is given by:
\begin{align}
    \rho_\mathrm{dm}(a) &=
    a^{-3} \big( \rho_\mathrm{dm,0} + 3\rho_\mathrm{de,0} I(a) \big),
    \label{eq:rho_dm_1} \\
    I(a) &=
    \int_1^a (\delta + \eta \tilde{a}) \tilde{a}^{-1 -3(w_0 + w_a + \delta)} e^{3(\eta - w_a) (1 - \tilde{a})} \,\mathrm{d}\tilde{a},
    \label{eq:integral}
\end{align}
with $\rho_\mathrm{dm,0}$ being the present value of the DM energy density. The first term in Eq.~\eqref{eq:rho_dm_1} recovers the standard scaling of non-interacting pressureless DM. All information regarding the interaction is encapsulated in the second term, which acts as a cumulative ``memory'' of the dark sector interaction history.

The integral $I(a)$ can be solved analytically in terms of confluent hypergeometric functions ${}_1F_1$. Defining the parameters $p = -1 - 3(w_0 + w_a + \delta)$ and $k = 3(\eta - w_a)$, the solution reads (see Appendix~\ref{sec:derivation_rho_de} for details):
\begin{equation}
    I(a) =
    e^k \left[ \frac{\delta}{p + 1} J(p + 1, a) + \frac{\eta}{p + 2} J(p + 2, a) \right],
    \label{eq:integral_sol_1}
\end{equation}
with the function $J(s, a)$ defined as:
\begin{equation}
    J(s, a) =
    a^s {}_1F_1(s; s + 1; -ka) - {}_1F_1(s; s + 1; -k).
    \label{eq:integral_sol_2}
\end{equation}
Provided that $I(a)$ remains finite as $a \to 0$, the energy density of DM retains information of the interaction even in the early Universe, $\rho_\mathrm{dm} \propto [\rho_\mathrm{dm,0} + 3\rho_\mathrm{de,0} I(0)] a^{-3}$, effectively re-scaling the initial abundance of DM in regimes where DE is otherwise negligible.

\subsection{Stability and pathologies}

IDE models are known for containing unphysical pathologies; specifically, there might be regions of the $(w_0, w_a, \delta, \eta)$ parameter space that may result in undefined or imaginary energy densities. In the following, we separate the discussion between constraints required for a consistent background evolution, and those necessary for the theoretical stability of linear perturbation equations.

\subsubsection{Background viability}

Regarding DE, we first note that, provided that radiation and matter do not exhaust the energy content of the Universe---their current total energy density does not amount to the critical energy density $\rho_\mathrm{c} \!=\! 3H_0^2/(8 \pi G)$---the current energy density of DE remains positive, $\rho_\mathrm{de,0}>0$. It then follows from Eq.~\eqref{eq:rho_de} that $\rho_\mathrm{de}(a) > 0$ over all cosmic history, regardless of the direction of the energy flow between DM and DE. Nonetheless, we must ensure that the interaction does not alter the standard matter-radiation domination epochs, enforcing DE to be subdominant in the past. As the time-evolution of the physical DE is governed by $w_\mathrm{de}^\mathrm{eff}(a)$, the latter requires $w_\mathrm{de}(a) < 0$ at early times ($a \to 0$), leading to the condition
\begin{equation}
    w_0 + w_a + \delta < 0.
    \label{eq:early_time_constraint}
\end{equation}

While $\rho_\mathrm{de}(a)$ is strictly positive for $\rho_\mathrm{de,0} > 0$, the DM energy density in Eq.~\eqref{eq:rho_dm_1} involves a subtraction if the interaction term $I(a)$ becomes negative, which occurs when the time-dependent coupling $\delta + \eta a$ is positive.
We do not artificially restrict the parameter space to regions where $\rho_\mathrm{dm}$ is enforced to be positive. Nonetheless, in the subsequent observational analysis we will reject any parameter combination that leads to a negative squared Hubble parameter $H(z)^2 < 0$ at any point in the cosmic history. In addition, a well-defined DM requires preventing the hypergeometric terms in $\rho_\mathrm{dm}$ from becoming undefined or creating a polynomial divergence stronger than $a^{-3}$. The latter can be ensured by restricting the parameter space via the inequality~\eqref{eq:early_time_constraint}.

\subsubsection{Perturbative stability (the doom factor \textbf{d})}

While our study primarily focuses on the background evolution, the physical viability of IDE models is often constrained by the stability of linear perturbations. To perform such analysis, a phenomenological covariant specification of $Q^\nu$ is required. Following~\cite{Gavela:2009cy}, one can adopt the ansatz that the interaction vector is parallel to the DM four-velocity, $Q^\mu = Q u_\mathrm{dm}^\mu$, so that the evolution of density contrasts in the dark sector is governed by a characteristic parameter known as the ``doom factor'' $\mathbf{d}$. This quantity arises from the momentum transfer equation in the DE rest frame, and is defined as~\cite{Gavela:2009cy}:
\begin{equation}
    \mathbf{d} = \frac{Q}{3H \rho_\mathrm{de} (1 + w_\mathrm{de})}.
    \label{eq:doom_factor}
\end{equation}
The sign of $\mathbf{d}$ determines the fate of large-scale perturbations. If $\mathbf{d} > 0$, the interaction term acts as a driving force that overcomes the Hubble friction, leading to a non-adiabatic instability where the DE density contrast grows exponentially. To avoid this ``runaway'' regime and ensure stable cosmology consistent with structure formation, it is customary to impose the stability condition $\mathbf{d} < 0$.

For the specific interaction in Eq.~\eqref{eq:Q_linear}, the condition $\mathbf{d} < 0$ implies that the interaction strength $(\delta + \eta a)$ and the term $(1 + w_\mathrm{de})$ must have opposite signs. This leads to two distinct stability regimes:
\begin{itemize}
    \item \textbf{Quintessence regime ($w_\mathrm{de} > -1$):} Stability requires a negative interaction coupling (for which $Q < 0$), implying energy must flow from DM to DE:
    \begin{equation}
        w_0 > -1, \quad 
        w_0 + w_a > -1, \quad 
        \delta < 0, \quad 
        \delta + \eta < 0.
    \end{equation}

    \item \textbf{Phantom regime ($w_\mathrm{de} < -1$):} Stability requires a positive interaction coupling (so that $Q > 0$), implying energy must flow from DE to DM:
    \begin{equation}
        w_0 < -1, \quad 
        w_0 + w_a < -1, \quad 
        \delta > 0, \quad 
        \delta + \eta > 0.
    \end{equation}
\end{itemize}
Under these conditions, models that cross the phantom divide ($w_\mathrm{de} \!=\! -1$) or the zero-coupling line ($Q \!=\! 0$) during cosmic evolution are forbidden, as they yield instabilities in the perturbation equations~\cite{Gavela:2009cy}.

Since the primary scope of this work is an observational analysis relying exclusively on background expansion history, we do not explicitly implement the full perturbation equations. Instead, by invoking the Parametrized Post-Friedmann (PPF) mechanism~\cite{Hu:2008zd, Fang:2008sn, Li:2014eha} as a theoretical safeguard of large-scale stability, we relax the constraints imposed by the doom factor $\mathbf{d}$.%
\footnote{The PPF approach successfully regularizes both the phantom divide crossing ($w_\mathrm{de} \!=\! -1$) and the potential large-scale instabilities induced by dark sector interactions. We expect that extending established PPF results to our general linear interaction kernel~\eqref{eq:Q_linear} and the CPL parametrization will ensure a well-behaved perturbation sector. For a successful implementation of the PPF in the constant coupling and constant equation of state scenarios, see~\cite{Li:2014cee}.}
This allows us to fully explore the phenomenological parameter space---including dynamical crossings of the phantom divide and zero-coupling lines---imposing only the background viability conditions given by the inequality~\eqref{eq:early_time_constraint} and $H(z)^2 \geq 0$ at all times.

\section{\label{sec:methodology} Data and methodology}

\subsection{Datasets}

\subsubsection{Type Ia supernovae (Pantheon+)}

We rely on the Pantheon$+$ compilation~\cite{Scolnic:2021amr, Brout:2022vxf}, consisting on a set of $1701$ light curves of $1550$ distinct SN covering the redshift range $0.001 < z < 2.26$. To mitigate the impact of peculiar velocity uncertainties, which dominate at low redshifts, we restrict our analysis to the subset of SN at $z > 0.01$~\cite{Peterson:2021hel, Carr:2021lcj}.

The theoretical distance modulus of the $i$th SN is given by
\begin{equation}
    \mu_{\mathrm{th},i}(z_{\mathrm{hel},i}, z_{\mathrm{HD},i}; \boldsymbol{\theta}) =
    25 + 5\log_{10} d_\mathrm{L}(z_{\mathrm{hel},i}, z_{\mathrm{HD},i}; \boldsymbol{\theta}),
\end{equation}
where the luminosity distance $d_\mathrm{L}$, in units of Mpc, is defined as
\begin{equation}
    d_\mathrm{L}(z_{\mathrm{hel},i}, z_{\mathrm{HD},i}; \boldsymbol{\theta}) =
    (1 + z_{\mathrm{hel},i}) \int_0^{z_{\mathrm{HD},i}} \frac{\mathrm{d}z}{H(z; \boldsymbol{\theta})}.
\end{equation}
Here, $z_\mathrm{hel}$ denotes the heliocentric redshift, and $z_\mathrm{HD}$ is the Hubble diagram redshift, which accounts for peculiar velocity corrections~\cite{Carr:2021lcj}. $\boldsymbol{\theta}$ will denote the generic vector of cosmological parameters.

We compute the residuals between the observed peak magnitude $m_{\mathrm{obs},i}$ and the theoretical prediction $\Delta \mu_{\mathrm{th},i}$ as
\begin{equation}
    \Delta \mu_i = m_{\mathrm{obs},i} - M - \mu_{\mathrm{th},i},
    \label{eq:difference_distance_modulus}
\end{equation}
where $M$ is the fiducial absolute magnitude of SN. Since $M$ is degenerate with $H_0$, and we aim to remain agnostic regarding the Cepheid calibration from SH0ES to avoid potential tensions with the CMB~\cite{Brout:2022vxf}, we treat $M$ as a nuisance parameter. Then, assuming a flat prior on $M$, we analytically marginalize over this parameter. Following the methodology of Ref.~\cite{Conley_2010}, the marginalized $\chi_\mathrm{SN}^2$ is given by:
\begin{equation}
    \chi_\mathrm{SN}^2 =
    \Delta \boldsymbol{\mu}_{0}^T \cdot \boldsymbol{C}_\mathrm{SN}^{-1} \cdot \Delta \boldsymbol{\mu}_{0} - \frac{\beta_\mathrm{SN}^2}{\kappa_\mathrm{SN}} + \ln \left( \frac{\kappa_\mathrm{SN}}{2\pi} \right),
    \label{eq:chi2_SN}
\end{equation}
where $\Delta \boldsymbol{\mu}_{0} \equiv \boldsymbol{m}_{\mathrm{obs}} - \boldsymbol{\mu}_{\mathrm{th}}$ corresponds to the residual vector~\eqref{eq:difference_distance_modulus} but omitting the fiducial absolute magnitude $M$. The marginalization terms are defined as
\begin{equation}
    \kappa_\mathrm{SN} = \sum_{i,j} (\boldsymbol{C}_\mathrm{SN}^{-1})_{ij} , \qquad
    \beta_\mathrm{SN} = \sum_{i,j} (\boldsymbol{C}_\mathrm{SN}^{-1})_{ij} \Delta \mu_{0,j},
\end{equation}
where $\boldsymbol{C}_\mathrm{SN}$ represents the total covariance matrix (statistical plus systematic) provided by the Pantheon$+$ collaboration.

\subsubsection{Cosmic microwave background (compressed likelihood)}

To constrain the cosmological parameters at early times, we consider the shift parameters derived from the CMB power spectrum~\cite{Wang:2007mza}. This approach incorporates the essential geometric information of the CMB into a compact data vector, avoiding the computational expense of evaluating the full Boltzmann hierarchy at every step of the likelihood computation.

The use of this compressed likelihood is well-justified in our phenomenological framework. Because DE is enforced to be subdominant in the past, the relative contribution $Q/(H \rho_\mathrm{tot})$ of the interaction kernel~\eqref{eq:Q_linear} naturally vanishes during the radiation-dominated and recombination epochs. Consequently, the pre-recombination physics remains identical to the standard $\Lambda$CDM cosmology, allowing us to safely rely on geometric background observables. This ensures that the comoving sound horizon $r_\mathrm{s}(z; \boldsymbol{\theta})$, defined as
\begin{equation}
    r_\mathrm{s}(z; \boldsymbol{\theta}) =
    \int_z^\infty \frac{c_\mathrm{s}(z^\prime)\, \mathrm{d}z^\prime}{H(z^\prime; \boldsymbol{\theta})},
    \label{eq:sound_horizon}
\end{equation}
is unaffected by the dark sector coupling. Here, $c_\mathrm{s}(z)$ is the standard sound speed of the photon-baryon fluid:
\begin{equation}
    c_\mathrm{s}(z) =
    \frac{1}{\sqrt{3 \big( 1 + R_\mathrm{b}(z) \big)}}
    ,\quad
    R_\mathrm{b}(z) =
    \frac{3 \Omega_\mathrm{b}}{4 \Omega_\gamma} (1 + z)^{-1},
\end{equation}
where $\Omega_\gamma \approx 2.47 \times 10^{-5} h^{-2}$ is the current photon density parameter fixed by the CMB temperature $T_\mathrm{CMB} = 2.7255\, \mathrm{K}$, and $h \!=\! H_0/(100\, \mathrm{km} \, \mathrm{s}^{-1} \, \mathrm{Mpc}^{-1})$. We fix the fractional radiation density $\Omega_\mathrm{r}$ consistent with $T_\mathrm{CMB}$ and the effective number of neutrino degrees of freedom $N_\mathrm{eff} = 3.044$.

The geometric CMB probe is encapsulated by the shift parameter $R$ and the acoustic angular scale $\ell_a$, defined respectively as
\begin{align}
    R(\boldsymbol{\theta}) &= \sqrt{\Omega_\mathrm{m} H_0^2} \, D_\mathrm{M}(z_*; \boldsymbol{\theta}), \\
    \ell_a(\boldsymbol{\theta}) &= \pi \frac{D_\mathrm{M}(z_*; \boldsymbol{\theta})}{r_\mathrm{s}(z_*; \boldsymbol{\theta})},
\end{align}
where $D_\mathrm{M}(z; \boldsymbol{\theta})$ is the comoving angular diameter distance, 
\begin{equation}
    D_\mathrm{M}(z; \boldsymbol{\theta}) =
    \int_{0}^z \frac{\mathrm{d}z^\prime}{H(z^\prime; \boldsymbol{\theta})}, \label{eq:comoving_distance}
\end{equation}
and $z_*$ is the photon-decoupling redshift. To calculate the latter, we employ the fitting formula from Ref.~\cite{Aizpuru:2021vhd}:
\begin{equation}
    z_* = \frac{391.672\, \omega_\mathrm{m}^{-0.372296} + 937.422\, \omega_\mathrm{b}^{-0.97966}}{\omega_\mathrm{m}^{-0.0192951} \omega_\mathrm{b}^{-0.93681}} + \omega_\mathrm{m}^{-0.731631},
\end{equation}
where $\omega_i \equiv \Omega_i h^2$. This reproduces the exact numerical solution with a precision of $\sim 0.0005\%$.

The constraints are implemented via the following $\chi_\mathrm{CMB}^2$:
\begin{equation}
    \chi_\mathrm{CMB}^2 =
    \Delta \boldsymbol{v}_\mathrm{CMB}^T \cdot \boldsymbol{C}_\mathrm{CMB}^{-1} \cdot \Delta \boldsymbol{v}_\mathrm{CMB}.
    \label{eq:chi2_CMB}
\end{equation}
Here, $\Delta \boldsymbol{v}_\mathrm{CMB}$ represents the residual vector between the theoretical and observed values of the triad $(R, \ell_a, \Omega_\mathrm{b} h^2)$. Their mean values, as well as the corresponding covariance matrix $\boldsymbol{C}_\mathrm{CMB}$, are reported in Ref.~\cite{Bansal:2025ipo}, which are derived from a joint analysis of Planck PR3~\cite{Planck:2019nip} and ACT DR6~\cite{ACT:2023kun} data.

While the compressed CMB parameters provide robust constraints on the background expansion history $H(z)$ via the distance to the last scattering surface, it is important to state the caveats of this methodology. Bypassing the full perturbation equations makes this analysis insensitive to the late-time integrated Sachs-Wolfe effect and the polarization power spectrum, as well as to modifications to the growth rate of structures induced by the interaction kernel $Q$. Therefore, constraints derived in this work should be considered conservative, serving as a first step to evaluate the background viability. A complete treatment would require solving the fully-coupled perturbation equations to confront the models against the full CMB and large-scale structure power spectra.

\subsubsection{Baryon acoustic oscillations (DESI DR 2)}

We utilize the latest BAO measurements from the Data Release 2 (DR2) of the DESI~\cite{DESI:2025zgx}. This dataset provides constraints on the comoving distance $D_\mathrm{M}(z)/r_\mathrm{d}$, the Hubble distance $D_\mathrm{H}(z)/r_\mathrm{d}$, and the spherically averaged distance $D_\mathrm{V}(z)/r_\mathrm{d}$ across seven redshift bins spanning $0.295 \leq z_\mathrm{eff} \leq 2.33$. The theoretical prediction for the comoving distance $D_\mathrm{M}$ is given by Eq.~\eqref{eq:comoving_distance}, whereas that of the Hubble and spherically averaged distances is:
\begin{align}
    D_\mathrm{H}(z; \boldsymbol{\theta}) &=
    1/H(z; \boldsymbol{\theta}), \\
    D_\mathrm{V}(z; \boldsymbol{\theta}) &=
    {\left( z D_\mathrm{M}{(z; \boldsymbol{\theta})}^2 D_\mathrm{H}(z; \boldsymbol{\theta}) \right)}^{1/3}.
\end{align}
These distances must be normalized by the comoving sound horizon at the drag epoch, $r_\mathrm{d} \equiv r_\mathrm{s}(z_\mathrm{d}, \boldsymbol{\theta})$, evaluating the integral defined in Eq.~\eqref{eq:sound_horizon}. For the redshift at the drag epoch $z_\mathrm{d}$, we employ the high-precision fitting formula derived in~\cite{Aizpuru:2021vhd}, which is accurate to $\sim 0.001\%$:
\begin{equation}
    z_\mathrm{d} =
    \frac{1 + 428.169\, \omega_\mathrm{b}^{0.256459} \omega_\mathrm{m}^{0.616388} + 925.56\, \omega_\mathrm{m}^{0.751615}}{\omega_\mathrm{m}^{0.714129}}.
\end{equation}

Since BAO data are only sensitive to the $H_0 r_\mathrm{d}$ product, an external calibration of the sound horizon scale is required. This is achieved when performing joint analyses with the CMB due to its precise determination of the baryon abundance. However, in the absence of CMB data, we impose a Gaussian prior on the reduced fractional baryon density from Big Bang Nucleosynthesis (BBN)~\cite{Schoneberg:2024ifp} to break the degeneracy:
\begin{equation}
    \Omega_\mathrm{b} h^2 =
    0.02218 \pm 0.00055.
    \label{eq:BBN_omega_b}
\end{equation}

The $\chi_\mathrm{BAO}^2$ for this data is quantified by
\begin{equation}
    \chi_\mathrm{BAO}^2 = \Delta \boldsymbol{D}_\mathrm{BAO}^T \cdot \boldsymbol{C}_\mathrm{BAO}^{-1} \cdot \Delta \boldsymbol{D}_\mathrm{BAO},
\end{equation}
where $\Delta \boldsymbol{D}_\mathrm{BAO}$ is the vector of residuals between the observed and theoretical distance ratios. Following the DESI DR2 methodology~\cite{DESI:2025zgx}, we utilize the isotropic measurement $D_\mathrm{V}/r_\mathrm{d}$ at $z_\mathrm{eff} = 0.295$, while for the remainder effective redshifts we utilize the pairs $(D_\mathrm{M}/r_\mathrm{d}, D_\mathrm{H}/r_\mathrm{d})$. The covariance matrix $\boldsymbol{C}_\mathrm{BAO}$ includes both statistical and systematic correlations as provided by the DESI collaboration.

\subsection{Bayesian inference}

We perform a joint Bayesian analysis to constrain the free parameters of the IDE model. Given a set of observational data $\mathcal{D}$ and a theoretical model described by a parameter vector $\boldsymbol{\theta}$, the posterior probability distribution $\mathcal{P}(\boldsymbol{\theta} \vert \mathcal{D})$ is determined via Bayes' theorem:
\begin{equation}
    \mathcal{P}(\boldsymbol{\theta}|\mathcal{D}) = \frac{\mathcal{L}(\mathcal{D}|\boldsymbol{\theta}) \pi(\boldsymbol{\theta})}{\mathcal{Z}},
\end{equation}
where $\pi(\boldsymbol{\theta})$ represents the prior probability distribution, reflecting our knowledge of the parameters before considering the data $\mathcal{D}$. The likelihood $\mathcal{L}(\mathcal{D} \vert \boldsymbol{\theta})$ is constructed assuming Gaussian errors for the datasets described previously, such that $\mathcal{L} \propto \exp(-\chi^2/2)$, with $\chi^2 = \sum_i \chi_i^2$. The normalization constant, known as the Bayesian evidence $\mathcal{Z}$, is obtained by marginalizing the product of the likelihood and prior over the full parameter space:
\begin{equation}
    \mathcal{Z} = \int \mathcal{L}(\mathcal{D}|\boldsymbol{\theta}) \pi(\boldsymbol{\theta}) \, \mathrm{d}\boldsymbol{\theta}.
    \label{eq:evidence}
\end{equation}
For two competing models $\mathcal{M}_i$ and $\mathcal{M}_j$, the Bayes factor $\mathcal{B}_i^j = \mathcal{Z}_j/\mathcal{Z}_i$ quantifies the relative preference of the data for one model over the other. Specifically, following Jeffrey's empirical scale~\cite{Jeffreys61}: if $\ln \mathcal{B}_i^j < 1$, the support in favor of $\mathcal{M}_j$ is weak; when $1 < \ln \mathcal{B}_i^j < 2.5$, the evidence is considered substantial; for $2.5 < \ln \mathcal{B}_i^j < 5$, the support for $\mathcal{M}_j$ is strong, and if $\ln \mathcal{B}_i^j > 5$, the evidence in favor of $\mathcal{M}_j$ is regarded as decisive. It is worth emphasizing that this scale is not universal and could introduce potential biases depending on the context and classes of models~\cite{Nesseris:2012cq}.

To report consistent statistical claims throughout this work, our model comparison will only rely on Bayesian evidence $\mathcal{Z}$. Commonly employed statistics such as the maximum-a-posteriori $\Delta \chi_\mathrm{MAP}^2$, the Akaike Information Criteria~\cite{Akaike:1974xx} or the Bayesian Information Criteria~\cite{Schwarz:1978xx} are avoided: they only require information about the maximum likelihood achievable by a model, and, as opposed to $\mathcal{Z}$, these do not encode further information from the likelihood throughout all parameter space. Therefore, their applicability is limited to Gaussian or nearly-Gaussian posterior distributions~\cite{Liddle:2007fy}, which are not expected to hold true in extended models such as the IDE models.

We analyze a hierarchy of nine cosmological scenarios, spanning from the standard $\Lambda$CDM model to increasingly complex IDE models. These models are constructed systematically relaxing the constraints on the DE equation of state---progressing from a cosmological constant $\Lambda$ to $w_0$CDM (constant $w_\mathrm{de}$) and the CPL $w_0 w_a$CDM parametrization---and by introducing the interaction kernel from Eq.~\eqref{eq:Q_linear} with constant ($\delta$) or dynamical ($\delta, \eta$) coupling. Specifically, the models considered are:
\begin{itemize}
    \item[i)] \textbf{Non-interacting $\boldsymbol{(\delta = \eta = 0)}$:} $\Lambda$CDM, $w_0$CDM, and $w_0 w_a$CDM.
    \item[ii)] \textbf{Constant coupling $\boldsymbol{(\eta = 0)}$:} $\delta \Lambda$CDM, $\delta w_0$CDM, and $\delta w_0 w_a$CDM.
    \item[iii)] \textbf{Dynamical coupling:} $\delta \eta \Lambda$CDM, $\delta \eta w_0$CDM, and $\delta \eta w_0 w_a$CDM.
\end{itemize}
The most general case, $\delta \eta w_0 w_a$CDM, is described by the full parameter vector $\boldsymbol{\theta} = \{ \Omega_\mathrm{m}, H_0, \Omega_\mathrm{b}, w_0, w_a, \delta, \eta \}$, and all other scenarios are nested models recovered by fixing the appropriate parameters. The uniform priors adopted for these parameters are summarized in Table~\ref{tab:priors}. 

To test the impact of different observational probes on the dark sector coupling, we perform the inference over four distinct dataset combinations: CMB$+$BAO, CMB$+$SN, BAO$+$SN and CMB$+$BAO$+$SN (denoted as ``Full'' throughout Section~\ref{sec:results}). Results including BAO data in the absence of CMB incorporate the BBN prior on $\Omega_\mathrm{b} h^2$ to ensure calibration of the acoustic scale $r_\mathrm{d}$.

\begin{table}[b]
    \centering
    \setlength{\tabcolsep}{1em}
    \caption{Flat prior distributions for the sampled parameters over the interval $[x, y]$. As noted in the main text, these parameter ranges encompass regions of the parameter space that are both physical and nonphysical, ultimately requiring a correction of the Bayesian evidence obtained by the Nested Sampling algorithm \texttt{Nautilus} (see also Appendix~\ref{sec:evidence_correction}).}%
    \label{tab:priors}
    \begin{tabular}{cc}
        \hline\hline\rule{0pt}{.75em}%
        Parameter & Flat priors \\ \hline\rule{0pt}{1.em}%
        $\Omega_\mathrm{m}$ & $[0.01, 0.99]$ \\[0.2em]
        $H_0 $ & $[20, 100]$ \\[0.2em]
        $\Omega_\mathrm{b}$ & $[0.005, 0.1]$ \\[0.2em]
        $w_0$ & $[-3, 1]$ \\[0.2em]
        $w_a$ & $[-3, 2]$ \\[0.2em]
        $\delta$ & $[-2, 2]$ \\[0.2em]
        $\eta$ & $[-2, 2]$ \\ \hline\hline 
    \end{tabular}
\end{table}

\begin{figure*}[!t]
    \centering
    \includegraphics[scale=0.65]{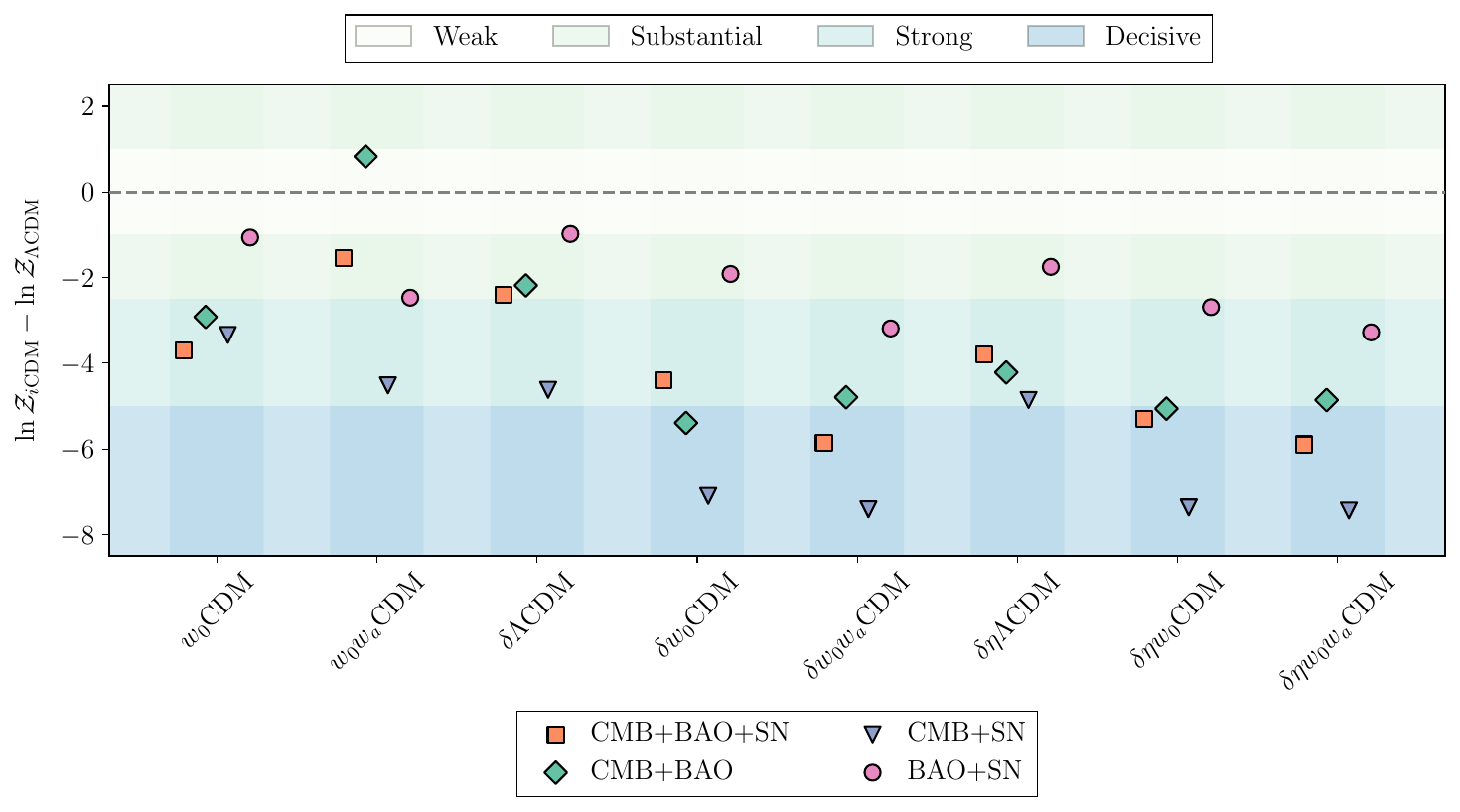}
    \caption{Comparison of the Bayes factors $\ln \mathcal{B}_{\Lambda\mathrm{CDM}}^{i\mathrm{CDM}} = \log \mathcal{Z}_{i \mathrm{CDM}} - \log \mathcal{Z}_{\Lambda \mathrm{CDM}}$, considering $\Lambda$CDM as the baseline reference, for different dataset combinations. $i$CDM stands for the different constant and dynamical coupling cosmologies, indicated in the horizontal axis. Positive values reveal evidence in favor of $i$CDM models, whereas negative values favor $\Lambda$CDM. The horizontal bands reflect the significance of the evidence according to Jeffreys' scale~\cite{Jeffreys61}; the alternating opacity is used to clearly group the dataset results belonging to each model.}
    \label{fig:evidence}
\end{figure*}

To explore the parameter space and compute the Bayesian evidence, we employ the \texttt{Nautilus} sampler~\cite{Lange:2023ydq}. \texttt{Nautilus} utilizes importance nested sampling boosted by deep learning; it trains neural networks to approximate the boundary of the iso-likelihood contours, significantly increasing sampling efficiency for complex, high-dimensional posteriors.
As the stopping criterion of the algorithm does not ensure proper convergence of the results, we repeated multiple runs varying the hyperparameters of \texttt{Nautilus}, mainly the number of live points $\texttt{n}_\texttt{live}$, and checked the results (posterior distributions and Bayesian evidences) remained stable.%
\footnote{As an additional test, we visualized the shell-bound occupation for all runs, and concluded that \texttt{Nautilus} correctly approximated the bounds $B_i$ that encompass the region of the parameter space where $\ln \mathcal{L} > \ln \mathcal{L}_i$ with $\ln \mathcal{L}_i$ increasing with $i$. The interested reader may find further details of this criterion in the following user-guide: \url{https://nautilus-sampler.readthedocs.io/en/latest/discussion/convergence.html}.}
The final results reported in this work were performed with the default hyperparameters of \texttt{Nautilus}, with the exception of the live point population, which is set to $\texttt{n}_\texttt{live} \!=\! 30.000$, and the exploration discard flag enabled ($\texttt{discard\_exploration} \!=\! \texttt{True}$), to ensure precise evidence and parameter constrain estimations. The priors in Table~\ref{tab:priors} encompass regions of the parameter space that do not satisfy the physical constraints given by the inequality~\eqref{eq:early_time_constraint} and $H(z)^2 > 0$. The procedure for enforcing these stability criteria and computing an unbiased estimate of the Bayesian evidence is detailed in Appendix~\ref{sec:evidence_correction}.

\section{\label{sec:results} Results and discussion}

In this section, we present the observational constraints and Bayesian model comparison for the nine cosmological scenarios outlined in Section~\ref{sec:methodology}. The marginalized posteriors---median and $68\%$ credible intervals---for the sampled parameters across the four dataset combinations are summarized in Tables~\ref{tab:results_non_interacting},~\ref{tab:results_constant_coupling}, and~\ref{tab:results_dynamical_coupling} for the non-interacting, constant coupling, and dynamical coupling models, respectively. Additionally, Figure~\ref{fig:evidence} provides a visual representation of the Bayes' factors $\ln \mathcal{Z}_{i\mathrm{CDM}} - \ln \mathcal{Z}_{\Lambda\mathrm{CDM}}$, using the $\Lambda$CDM model as the baseline reference model; positive values favor the extended models, whereas negative values support the $\Lambda$CDM model.

\begin{table*}[t]
    \centering
    \caption{Median and $68\%$ asymmetric credible intervals for the $\Lambda$CDM, $w_0$CDM and the $w_0 w_a$CDM models (CPL parametrization). Horizontal lines indicate that the corresponding parameter is absent for a model and is not sampled.}
    \label{tab:results_non_interacting}
    \begin{ruledtabular}
    \begin{tabular}{l l c c c c c c c}
        Model & Dataset & $\Omega_{\mathrm{m}}$ & $H_0$ & $100\Omega_{\mathrm{b}}$ & $w_0$ & $w_a$ \vspace{.1em} \\ \hline\rule{0pt}{1.1em}%
        $\Lambda$CDM & CMB+BAO & $0.3006_{-0.0037}^{+0.0038}$ & $68.40_{-0.29}^{+0.28}$ & $4.824_{-0.032}^{+0.032}$ & ------ & ------ \\[0.25em]
         & CMB+SN & $0.3191_{-0.0067}^{+0.0067}$ & $67.06_{-0.46}^{+0.47}$ & $4.967_{-0.053}^{+0.053}$ & ------ & ------ \\[0.25em]
         & BAO+SN & $0.3045_{-0.0079}^{+0.0080}$ & $68.71_{-0.58}^{+0.60}$ & $4.696_{-0.077}^{+0.079}$ & ------ & ------ \\[0.25em]
         & Full & $0.3019_{-0.0037}^{+0.0036}$ & $68.30_{-0.28}^{+0.28}$ & $4.833_{-0.032}^{+0.031}$ & ------ & ------ \\[1.em]
        $w_0$CDM & CMB+BAO & $0.2938_{-0.0075}^{+0.0074}$ & $69.33_{-0.93}^{+0.97}$ & $4.68_{-0.14}^{+0.14}$ & $-1.040_{-0.039}^{+0.037}$ & ------ \\[0.25em]
         & CMB+SN & $0.3240_{-0.0087}^{+0.0088}$ & $66.46_{-0.79}^{+0.81}$ & $5.06_{-0.12}^{+0.12}$ & $-0.975_{-0.028}^{+0.029}$ & ------ \\[0.25em]
         & BAO+SN & $0.2974_{-0.0087}^{+0.0084}$ & $66.3_{-1.3}^{+1.2}$ & $5.05_{-0.18}^{+0.20}$ & $-0.912_{-0.040}^{+0.039}$ & ------ \\[0.25em]
         & Full & $0.3043_{-0.0051}^{+0.0049}$ & $67.95_{-0.56}^{+0.59}$ & $4.887_{-0.087}^{+0.088}$ & $-0.985_{-0.023}^{+0.023}$ & ------ \\[1.em]
        $w_0w_a$CDM & CMB+BAO & $0.356_{-0.022}^{+0.021}$ & $63.3_{-1.7}^{+2.0}$ & $5.58_{-0.33}^{+0.32}$ & $-0.38_{-0.22}^{+0.21}$ & $-1.86_{-0.62}^{+0.63}$ \\[0.25em]
         & CMB+SN & $0.319_{-0.012}^{+0.015}$ & $67.0_{-1.4}^{+1.3}$ & $4.98_{-0.19}^{+0.22}$ & $-0.92_{-0.11}^{+0.11}$ & $-0.27_{-0.56}^{+0.56}$ \\[0.25em]
         & BAO+SN & $0.303_{-0.023}^{+0.015}$ & $66.8_{-2.5}^{+1.7}$ & $4.97_{-0.25}^{+0.38}$ & $-0.890_{-0.055}^{+0.062}$ & $-0.17_{-0.45}^{+0.47}$ \\[0.25em]
         & Full & $0.3108_{-0.0056}^{+0.0056}$ & $67.59_{-0.57}^{+0.60}$ & $4.914_{-0.089}^{+0.087}$ & $-0.839_{-0.054}^{+0.055}$ & $-0.60_{-0.22}^{+0.21}$ \\
    \end{tabular}
    \end{ruledtabular}
\end{table*}

\subsection{Non-interacting models}

We first validate our analysis pipeline by examining the non-interacting models ($\Lambda$CDM, $w_0$CDM, and $w_0 w_a$CDM). Overall, the parameter constraints are in excellent agreement with the official results reported by the DESI collaboration in their DR2~\cite{DESI:2025zgx}.

For the baseline $\Lambda$CDM model, we observe a minor $\sim\! 1\sigma$ shift in the $\Omega_\mathrm{m}$ and $H_0$ constraints when using the CMB$+$BAO data combination, compared to the full perturbation analyses in the literature. This slight deviation is a consequence of employing the compressed CMB likelihood rather than the full temperature and polarization power spectra. The compression encapsulates the background geometry but loses sensitivity to late-time perturbations, such as the integrated Sachs-Wolfe effect. When extending the parameter space to include dynamical DE parameters $w_0$ and $w_a$, our constraints remain highly consistent with the DESI DR2 findings.

Regarding model comparison,  Bayesian evidence penalizes the additional degrees of freedom introduced by the evolving DE models. As illustrated in Figure~\ref{fig:evidence}, we find no statistical preference for $w_0$CDM or $w_0 w_a$CDM over $\Lambda$CDM. The only exception is the CMB$+$DESI joint analysis, which yields \emph{weak} evidence ($\vert \ln \mathcal{B} \vert < 1$) in favor of the CPL parametrization. For all other dataset combinations, the evidence ranges from substantial to strong in favor of the $\Lambda$CDM baseline. While maximum-likelihood criteria might highlight a preference for evolving DE---as noted in several recent frequentist analyses---the Bayesian evidence calculation confirms that current geometric data do not justify robustly the increased model complexity in the non-interacting dark sector, in agreement with Refs.~\cite{Ong:2025utx, Ong:2026tta}.

\subsection{Constant coupling models}

\begin{table*}[ht]
    \centering
    \caption{Constraints for the constant coupling interaction models, where $Q = 3H \delta \rho_\mathrm{de}$, when considering $w_\mathrm{de} = -1$ ($\delta \Lambda$CDM model), $w_\mathrm{de} = w_0$ ($\delta w_0$CDM), and $w_\mathrm{de} = w_0 + w_a(1-a)$ ($\delta w_0 w_a$CDM model). Horizontal lines indicate that the corresponding parameter is absent for a model and is not sampled.}
    \label{tab:results_constant_coupling}
    \begin{ruledtabular}
    \begin{tabular}{l l c c c c c c c}
        Model & Dataset & $\Omega_{\mathrm{m}}$ & $H_0$ & $100\Omega_{\mathrm{b}}$ & $\delta$ & $w_0$ & $w_a$ \vspace{.1em} \\ \hline\rule{0pt}{1.1em}%
        $\delta\Lambda$CDM & CMB+BAO & $0.3001_{-0.0037}^{+0.0039}$ & $68.56_{-0.29}^{+0.30}$ & $4.758_{-0.039}^{+0.040}$ & $0.0064_{-0.0023}^{+0.0022}$ & ------ & ------ \\[0.25em]
         & CMB+SN & $0.332_{-0.016}^{+0.017}$ & $66.1_{-1.2}^{+1.2}$ & $5.12_{-0.19}^{+0.18}$ & $-0.0062_{-0.0082}^{+0.0071}$ & ------ & ------ \\[0.25em]
         & BAO+SN & $0.357_{-0.026}^{+0.026}$ & $66.5_{-1.2}^{+1.2}$ & $5.01_{-0.18}^{+0.19}$ & $0.084_{-0.040}^{+0.040}$ & ------ & ------ \\[0.25em]
         & Full & $0.3016_{-0.0037}^{+0.0037}$ & $68.44_{-0.29}^{+0.29}$ & $4.771_{-0.040}^{+0.040}$ & $0.0062_{-0.0023}^{+0.0022}$ & ------ & ------ \\[1.em]
        $\delta w_0$CDM & CMB+BAO & $0.3063_{-0.0087}^{+0.0087}$ & $67.7_{-1.0}^{+1.1}$ & $4.88_{-0.15}^{+0.16}$ & $0.0077_{-0.0029}^{+0.0029}$ & $-0.965_{-0.046}^{+0.044}$ & ------ \\[0.25em]
         & CMB+SN & $0.324_{-0.044}^{+0.038}$ & $66.4_{-1.9}^{+2.4}$ & $5.07_{-0.36}^{+0.30}$ & $0.000_{-0.034}^{+0.025}$ & $-0.97_{-0.12}^{+0.12}$ & ------ \\[0.25em]
         & BAO+SN & $0.53_{-0.31}^{+0.19}$ & $66.7_{-1.3}^{+1.4}$ & $4.98_{-0.19}^{+0.21}$ & $0.46_{-0.54}^{+0.92}$ & $-1.37_{-0.93}^{+0.54}$ & ------ \\[0.25em]
         & Full & $0.3097_{-0.0054}^{+0.0054}$ & $67.30_{-0.61}^{+0.63}$ & $4.939_{-0.092}^{+0.093}$ & $0.0085_{-0.0025}^{+0.0025}$ & $-0.946_{-0.027}^{+0.026}$ & ------ \\[1.em]
        $\delta w_0 w_a$CDM & CMB+BAO & $0.346_{-0.025}^{+0.025}$ & $64.1_{-2.1}^{+2.3}$ & $5.44_{-0.37}^{+0.37}$ & $0.0025_{-0.0033}^{+0.0038}$ & $-0.51_{-0.27}^{+0.27}$ & $-1.41_{-0.86}^{+0.84}$ \\[0.25em]
         & CMB+SN & $0.361_{-0.042}^{+0.032}$ & $65.1_{-1.6}^{+2.1}$ & $5.28_{-0.32}^{+0.27}$ & $-0.045_{-0.048}^{+0.046}$ & $-0.95_{-0.13}^{+0.12}$ & $-1.1_{-1.2}^{+1.1}$ \\[0.25em]
         & BAO+SN & $0.56_{-0.30}^{+0.17}$ & $67.4_{-2.2}^{+1.7}$ & $4.87_{-0.24}^{+0.33}$ & $0.53_{-0.58}^{+0.89}$ & $-1.39_{-0.87}^{+0.56}$ & $-0.25_{-0.47}^{+0.48}$ \\[0.25em]
         & Full & $0.3112_{-0.0055}^{+0.0056}$ & $67.35_{-0.60}^{+0.59}$ & $4.933_{-0.090}^{+0.091}$ & $0.0059_{-0.0032}^{+0.0034}$ & $-0.889_{-0.057}^{+0.060}$ & $-0.27_{-0.27}^{+0.25}$ \\
    \end{tabular}
    \end{ruledtabular}
\end{table*}

\begin{figure}[ht]
    \centering
    \includegraphics{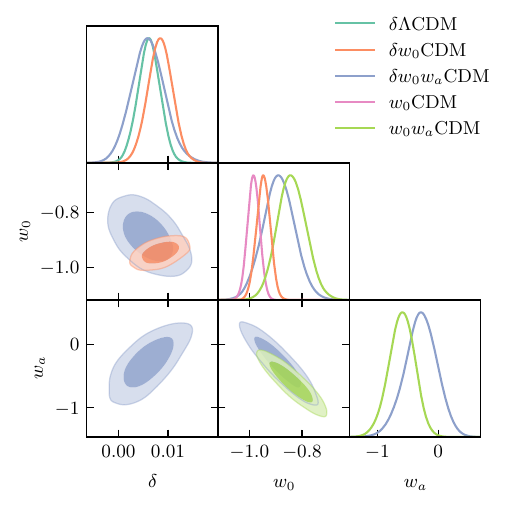}
    \caption{One- and two-dimensional marginalized posterior distributions (CMB$+$BAO$+$SN) of the dark sector parameters $(\delta, w_0, w_a)$ for $\eta = 0$ models. We include the posterior distributions of $w_0$ and $w_a$ for the $w_0$CDM and $w_0 w_a$CDM (CPL parametrization) models as a reference.}
    \label{fig:constant_coupling}
\end{figure}

We now consider the extension of the previous models by introducing a constant interaction coupling $\delta$ in the dark sector. The constraints for the $\delta \Lambda$CDM, $\delta w_0$CDM, and $\delta w_0 w_a$CDM scenarios are reported in Table~\ref{tab:results_constant_coupling}. To clearly visualize the correlations within the dark sector, the one- and two-dimensional marginalized posterior distributions for $\{ \delta, w_0, w_a \}$ using the CMB+BAO+SN combination are displayed in Figure~\ref{fig:constant_coupling}.

In the absence of the compressed CMB likelihood, we find that the constraints on the matter density $\Omega_\mathrm{m}$ severely degrade when the DE equation-of-state parameters are allowed to vary. Specifically, the one-dimensional constraints widen significantly to $\Omega_\mathrm{m} = 0.53_{-0.31}^{+0.19}$ and $\Omega_\mathrm{m} = 0.56_{-0.30}^{+0.17}$ for the $\delta w_0$CDM and $\delta w_0 w_a$CDM models, respectively. In the two-dimensional posteriors, we observed a strong anti-correlation between $\delta$ and $w_0$, and consequently, a large degeneracy in both the $(\Omega_\mathrm{m}, \delta)$ and $(\Omega_\mathrm{m}, w_0)$ planes. A similar trend was recently observed in Ref.~\cite{Figueruelo:2026eis}, even when incorporating cosmic chronometer measurements. Despite these degeneracies, the constraints on the Hubble constant $H_0$ remain robust across all three models.

When incorporating the CMB data, this late-time degeneracy is broken. The CMB precisely measures the distance to the last scattering surface, and for spatially flat cosmologies, it presents a well-known degeneracy where $\Omega_\mathrm{m} h^3$ is approximately constant (see e.g.~\cite{Howlett_2012, 2dFGRSTeam:2002tzq}). Because $H_0$ is well-calibrated by the low-redshift probes, the total matter density $\Omega_\mathrm{m}$ becomes highly constrained. This restricts the allowed energy transfer history, breaking the $(\delta, w_0)$ degeneracy and enabling a much sharper determination of the coupling parameter $\delta$.

Interestingly, the marginalized posteriors mildly support a non-vanishing interaction when the CMB is included. While the CMB$+$SN combination yields a $\delta$ compatible with zero within $\sim\! 1\sigma$ (with a slight preference for energy flow from DM to DE, $\delta < 0$), the inclusion of BAO data (CMB$+$BAO and CMB$+$BAO$+$SN) shifts the preference toward a positive coupling. By computing the probability mass enclosed by the highest posterior density contours for CMB$+$BAO$+$SN, we find that the $\Lambda$ point ($w_\mathrm{de} = -1$, $w_a = \delta = \eta = 0$) is excluded at $2.7\sigma$, $2.8\sigma$, and $2.9\sigma$ in the joint dark-sector parameter spaces of the $\delta \Lambda$CDM, $\delta w_0$CDM, and $\delta w_0 w_a$CDM models, respectively.

However, this apparent preference for an interacting dark sector vanishes when evaluated through the lens of Bayesian model comparison: the Bayes' factors severely penalize the expanded prior volumes of these additional parameters. As shown in Figure~\ref{fig:evidence}, the inclusion of new degrees of freedom (moving from $\delta \Lambda$CDM to $\delta w_0 w_a$CDM) systematically decreases the evidence $\mathcal{Z}$ of these models, heavily favoring the standard $\Lambda$CDM cosmology. Specifically, while the $\delta \Lambda$CDM model is only substantially or strongly penalized depending on the dataset, the full $\delta w_0 w_a$CDM faces either strong or decisive evidence against it for all joint data combinations.

Finally, we note that the joint analyses including the compressed CMB data consistently yield the strongest evidence in favor of $\Lambda$CDM. As stated before, the compressed likelihood only utilizes the shift parameters, and it is insensitive to late-time perturbation effects. A full Boltzmann analysis of the CMB temperature and polarization power spectra would likely penalize the IDE differently, and thus the results of the implementation of the CMB here can be considered as a preliminary baseline.

\subsection{Dynamical coupling models}

\begin{table*}[ht]
    \centering
    \caption{Median and $68\%$ confidence intervals for the dynamical coupling interaction models $Q = 3H(\delta + \eta a) \rho_\mathrm{de}$. Here, the $\delta \eta \Lambda$CDM corresponds to $w_\mathrm{de} = -1$, $\delta \eta w_0$CDM to $w_\mathrm{de} = w_0$, and $\delta \eta w_0 w_a$CDM to $w_0 = w_0 + w_a(1-a)$. Horizontal lines indicate that the corresponding parameter is absent for a model and is not sampled.}
    \label{tab:results_dynamical_coupling}
    \begin{ruledtabular}
    \begin{tabular}{l l c c c c c c c}
        Model & Dataset & $\Omega_{\mathrm{m}}$ & $H_0$ & $100\Omega_{\mathrm{b}}$ & $\delta$ & $\eta$ & $w_0$ & $w_a$ \vspace{.1em} \\ \hline\rule{0pt}{1.1em}%
        $\delta\eta\Lambda$CDM & CMB+BAO & $0.3017_{-0.0054}^{+0.0054}$ & $68.29_{-0.75}^{+0.72}$ & $4.80_{-0.10}^{+0.10}$ & $0.06_{-0.15}^{+0.13}$ & $-0.08_{-0.17}^{+0.19}$ & ------ & ------ \\[0.25em]
         & CMB+SN & $0.347_{-0.031}^{+0.024}$ & $65.5_{-1.2}^{+1.3}$ & $5.21_{-0.20}^{+0.20}$ & $-0.50_{-0.76}^{+0.76}$ & $0.61_{-0.96}^{+0.92}$ & ------ & ------ \\[0.25em]
         & BAO+SN & $0.383_{-0.052}^{+0.050}$ & $67.3_{-2.1}^{+1.7}$ & $4.89_{-0.23}^{+0.32}$ & $-0.15_{-0.42}^{+0.40}$ & $0.35_{-0.60}^{+0.63}$ & ------ & ------ \\[0.25em]
         & Full & $0.3053_{-0.0047}^{+0.0047}$ & $67.75_{-0.61}^{+0.60}$ & $4.873_{-0.087}^{+0.092}$ & $0.15_{-0.11}^{+0.11}$ & $-0.20_{-0.14}^{+0.15}$ & ------ & ------ \\[1.em]
        $\delta\eta w_0$CDM & CMB+BAO & $0.327_{-0.016}^{+0.013}$ & $65.9_{-1.3}^{+1.6}$ & $5.15_{-0.24}^{+0.21}$ & $-0.81_{-0.42}^{+0.56}$ & $1.13_{-0.78}^{+0.60}$ & $-0.73_{-0.16}^{+0.11}$ & ------ \\[0.25em]
         & CMB+SN & $0.340_{-0.042}^{+0.036}$ & $65.8_{-1.7}^{+2.1}$ & $5.16_{-0.32}^{+0.28}$ & $-0.53_{-0.74}^{+0.76}$ & $0.66_{-0.97}^{+0.91}$ & $-0.97_{-0.13}^{+0.11}$ & ------ \\[0.25em]
         & BAO+SN & $0.55_{-0.31}^{+0.19}$ & $67.6_{-2.0}^{+1.7}$ & $4.85_{-0.24}^{+0.31}$ & $0.24_{-0.72}^{+0.91}$ & $0.39_{-0.58}^{+0.62}$ & $-1.35_{-0.92}^{+0.55}$ & ------ \\[0.25em]
         & Full & $0.3113_{-0.0057}^{+0.0057}$ & $67.34_{-0.59}^{+0.61}$ & $4.931_{-0.090}^{+0.093}$ & $-0.29_{-0.30}^{+0.27}$ & $0.40_{-0.36}^{+0.40}$ & $-0.888_{-0.059}^{+0.059}$ & ------ \\[1.em]
        $\delta\eta w_0 w_a$CDM & CMB+BAO & $0.344_{-0.023}^{+0.023}$ & $64.3_{-1.9}^{+2.1}$ & $5.41_{-0.34}^{+0.34}$ & $-0.01_{-0.92}^{+0.98}$ & $0.0_{-1.3}^{+1.3}$ & $-0.52_{-0.25}^{+0.24}$ & $-1.44_{-0.95}^{+1.14}$ \\[0.25em]
         & CMB+SN & $0.364_{-0.039}^{+0.032}$ & $65.0_{-1.6}^{+2.0}$ & $5.30_{-0.32}^{+0.26}$ & $-0.24_{-0.95}^{+1.11}$ & $0.2_{-1.4}^{+1.2}$ & $-0.95_{-0.13}^{+0.12}$ & $-1.2_{-1.1}^{+1.3}$ \\[0.25em]
         & BAO+SN & $0.55_{-0.29}^{+0.18}$ & $67.6_{-2.0}^{+1.7}$ & $4.85_{-0.23}^{+0.30}$ & $0.43_{-0.95}^{+0.97}$ & $0.3_{-1.3}^{+1.0}$ & $-1.35_{-0.85}^{+0.53}$ & $-0.07_{-0.99}^{+0.97}$ \\[0.25em]
         & Full & $0.3112_{-0.0054}^{+0.0058}$ & $67.36_{-0.61}^{+0.61}$ & $4.932_{-0.092}^{+0.091}$ & $0.00_{-0.76}^{+1.03}$ & $0.0_{-1.3}^{+1.1}$ & $-0.880_{-0.060}^{+0.061}$ & $-0.29_{-0.95}^{+0.90}$ \\
    \end{tabular}
    \end{ruledtabular}
\end{table*}

\begin{figure}[ht]
    \centering
    \includegraphics{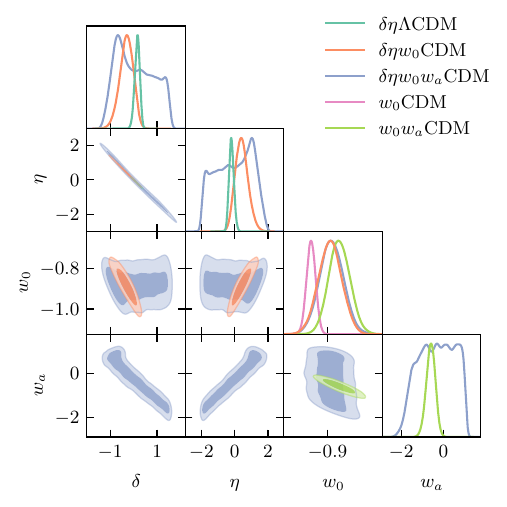}
    \caption{One- and two-dimensional marginalized posterior distributions of the dark sector parameters for the IDE models with dynamical coupling (CMB+BAO+SN data combination). For reference, we also include the posterior distributions of $w_0$ and $w_a$ for the $w_0$CDM and $w_0 w_a$CDM models.}
    \label{fig:dynamical_coupling}
\end{figure}

Lastly, we investigate the most general extension of our model hierarchy by promoting the interaction coupling to a linear function of the scale factor, governed by the additional parameter $\eta$. The constraints for the three dynamical scenarios---labeled as $\delta \eta \Lambda$CDM, $\delta \eta w_0$CDM, and $\delta \eta w_0 w_a$CDM---are summarized in Table~\ref{tab:results_dynamical_coupling}. To illustrate the degeneracies within the extended dark sector, the one- and two-dimensional marginalized posterior distributions for the parameters $\{\delta, \eta, w_0, w_a\}$ are presented in Figure~\ref{fig:dynamical_coupling}.

As anticipated from the constant coupling analysis, these extended models suffer from severe degeneracies when the CMB anchor is excluded. Relying solely on the late-time expansion history from BAO and SN, the dark sector becomes overly flexible. The interplay between the equation-of-state parameters $(w_0, w_a)$ and the dynamical coupling $(\delta, \eta)$ completely messes the constraints on the current matter density $\Omega_\mathrm{m}$, rendering its determination highly inaccurate. Although the Hubble constant $H_0$ remains remarkably stable and competitive across all models, the distance-redshift of late-time data cannot resolve the underlying composition of the Universe.

The inclusion of the compressed CMB shift parameters restores the constraints on $\Omega_\mathrm{m}$ and breaks the degeneracies with $w_0$ and $w_a$. However, an intrinsic degeneracy remains within the interaction sector itself, observed in the strong anti-correlation between $\delta$ and $\eta$. To preserve the integrated energy transfer $I(a)$ required to fit the data, an increase in the present-day coupling $\delta$ must be compensated by a negative $\eta$, ensuring the effective interaction during the matter-dominated era remains viable. Consequently, in the most general $\delta \eta w_0 w_a$CDM model, the dark sector parameter space remains largely unconstrained. By evaluating the joint probability mass enclosed by the highest posterior density regions in the dark sector parameter space for the CMB$+$BAO$+$SN data combination, we find that the standard $\Lambda$ point lies well within the $1.3\sigma$ and $1.5\sigma$ credible regions for all three models, indicating no significant deviation from standard cosmology.

This lack of parametric preference is significantly penalized according to the Bayesian evidence. The statistical trend mirrors and amplifies the constant coupling case: as the complexity of the dark sector increases, the evidence in favor of $\Lambda$CDM becomes ubiquitous. Specifically, for the $\delta \eta \Lambda$CDM model, all dataset combinations (except for BAO$+$SN) show a strong preference for the consensus model. When extending the equation of state of DE to $\delta \eta w_0$CDM and $\delta \eta w_0 w_a$CDM, the evidence against the interacting models becomes decisive ($\vert \ln \mathcal{B} \vert > 5$) for nearly all joint combinations involving the CMB.

Therefore, while dynamical IDE models possess the flexibility to fit current geometric and background expansion data, this large parameter space is statistically unjustified. Our Bayesian analysis renders the uncoupled, cosmological constant-driven Universe as perfectly consistent with the latest measurements from DESI BAO, Pantheon$+$ and a compressed background CMB data.

\section{\label{sec:conclusions} Conclusions}

In this work, we derived the analytical solutions for the background energy densities of dark matter and dark energy in an extended cosmological scenario. Specifically, we investigated a background interaction kernel proportional to the dark energy density with a dynamical coupling, $Q = 3H(\delta + \eta a) \rho_\mathrm{de}$, combined with a dynamical dark energy equation-of-state parameter characterized by the CPL parametrization, $w_\mathrm{de}(a) = w_0 + w_a(1 - a)$. We established constraints on the dark-sector parameter space $(\delta, \eta, w_0, w_a)$ to ensure the standard subdominance of dark energy at early times, which simultaneously guaranteed a well-defined evolution of the dark matter energy density throughout cosmic time. Beyond the background dynamics, we promoted the interaction to a phenomenological covariant four-vector $Q^\nu$ parallel to the dark matter four-velocity, allowing us to impose additional bounds required for perturbative stability. In particular, we followed the prescription of the doom factor $\mathbf{d}$~\cite{Gavela:2009cy} to prevent the exponential runaway growth of the dark energy density contrast.

Building on this theoretical foundation, we conducted an extensive observational analysis across a hierarchy of nine cosmological models. These were systematically categorized into non-interacting ($\Lambda$CDM, $w_0$CDM, and $w_0 w_a$CDM), constant coupling ($\delta \Lambda$CDM, $\delta w_0$CDM, and $\delta w_0 w_a$CDM), and dynamical coupling ($\delta \eta \Lambda$CDM, $\delta \eta w_0$CDM, and $\delta \eta w_0 w_a$CDM) cosmologies. To constrain the parameter space, we evaluated various joint combinations of geometric probes: the Pantheon$+$ Type Ia supernovae compilation, BAO from DESI Data Release 2, and the compressed CMB shift parameters from Planck and ACT. Furthermore, rather than restricting the sampling strictly to the stable regimes dictated by the doom factor $\mathbf{d}$, we invoked the Parametrized Post-Friedmann formalism as a theoretical guarantee for large-scale perturbative stability. This approach allowed for filtering only for fundamental background viability: well-defined expansion rate ($H(z)^2 \!>\! 0$) and proper early-time subdominance of dark energy.

Excluding the early-Universe CMB acoustic scale calibration leads to severe parametric degeneracies when a non-zero dark sector interaction is simultaneously combined with a $w_\mathrm{de} \!\neq\! -1$ equation-of-state parameter. Within the extended $(\delta, \eta, w_0, w_a)$, this interplay effectively reduces the constraining power of late-time geometric probes, resulting in a highly unrestricted determination of the present-time matter density $\Omega_\mathrm{m}$. Incorporating the compressed CMB likelihood breaks these degeneracies for most of the considered cosmologies by anchoring the high-redshift expansion history: when coupled with robust $H_0$ constraints provided by the late-time data, the total matter density becomes tightly determined. However, in the most general phenomenological extension, the $\delta \eta w_0 w_a$CDM model, the flexibility of the dark sector leaves the interacting parameters largely unconstrained across all dataset combinations.

For the constant coupling scenarios, the joint analysis of all datasets (CMB$+$BAO$+$SN) revealed a mild parametric deviation from the standard cosmological constant ($w_0 = -1$, $w_a = 0$, $\delta = 0$), lying in the $2.7\sigma$ to $2.9\sigma$ range. Nonetheless, when expanding the parameter space to include a dynamical coupling, the parametric preference is significantly reduced, with the deviation dropping to $1.3\sigma$ to $1.5\sigma$. In any case, this parameter-level behavior is replaced by a rigorous Bayesian model comparison. Utilizing the Bayesian evidence $\mathcal{Z}$, we found that the increased complexity of IDE models is universally penalized. Across all data combination, the Bayes factors consistently favored the $\Lambda$CDM baseline, yielding substantial ($\vert \ln \mathcal{B} \vert > 2$) to decisive ($\vert \ln \mathcal{B} \vert > 5$) evidence against the extended dark sector scenarios.

In conclusion, when evaluated against the current background expansion and geometric data, we find no statistical support for IDE models governed by $Q = 3H(\delta + \eta a) \rho_\mathrm{de}$ over the standard $\Lambda$CDM paradigm. While the present study maps the background dynamics, a complete analysis incorporating the full Boltzmann hierarchy lies beyond the scope of this work, though such a stringent analysis could further constrain the allowed parameter space and amplify (or alleviate) the statistical penalties reported here.

Finally, our findings underscore the value of complementing standard frequentist maximum-likelihood analyses with full Bayesian model comparison~\cite{Jeffreys61, Robert_2009}. Extended phenomenological models often introduce complex parameter degeneracies and enlarged prior volumes, for which Bayesian evidence provides a more clear trace. Nonetheless, the strong prior-dependence of the evidence, as well as the heuristic nature of the Jeffreys scale, require caution when interpreting the reported statistical results~\cite{Nesseris:2012cq, SolaPeracaula:2018wwm, Patel:2024odo}. To address this, the Frequentist-Bayesian method, which treats the Bayes factor as a frequentist random variable to which a $p$-value can be assigned, has recently been proposed to ensure reliability of model selection claims~\cite{Jenkins_2011, Keeley:2021dmx, Amendola:2024prl, Sakr:2025chr, Gonzalez-Fuentes:2026rgu}. Ultimately, we suggest that reporting Bayes factors alongside goodness-of-fit metrics offers the community a more comprehensive and transparent assessment of whether new observational data truly justify a departure from the $\Lambda$CDM baseline.

\begin{acknowledgments}
M.A.~acknowledges support from the Basque Government Grant No.~PRE\_2025\_2\_0301. M.A.~and R.L.~are supported by the Basque Government Grant IT1628-22, and by Grant PID2021-123226NB-I00 (funded by MCIN/AEI/10.13039/501100011033, by ``ERDF A way of making Europe''). This article is based upon work from COST Actions CosmoVerse CA21136 and CaLISTA CA21109, supported by COST (European Cooperation in Science and Technology). Computational resources were provided by the Solaris cluster at the University of the Basque Country (EHU).
\end{acknowledgments}


\appendix
\section{\label{sec:derivation_rho_de} \texorpdfstring{Derivation of the \\energy density of dark matter}{Derivation of the energy density of dark matter}}

In this Appendix, we detail the evaluation of the integral $I(a)$ introduced in Eq.~\eqref{eq:integral}. Defining the auxiliary parameters $k = 3(\eta - w_a)$ and $p = -1 - 3(w_0 + w_a + \delta)$, we have
\begin{equation}
    I(a) = e^k \int_1^a (\delta \tilde{a}^p + \eta \tilde{a}^{p+1}) e^{-k\tilde{a}} \, \mathrm{d}\tilde{a}.
\end{equation}
The problem reduces to evaluating the integral of the form $\int t^s e^{-kt} \mathrm{d}t$. Now, we can solve this by Taylor expanding the exponential function and noting that, since the series converges uniformly, we can interchange the summation and the integral:
\begin{equation}
    \int t^s e^{-k t}\, \mathrm{d}t =
    t^{s+1} \sum_{n = 0}^\infty \frac{(-kt)^n}{n! (s+n+1)}.
\end{equation}
We can rewrite the denominator using the properties of the Pocchammer symbol $a^{(n)} = a (a + 1) \dots (a + n  - 1)$. Noting that $s + n + 1 = (s + 1) (s + 2)^{(n)} / (s + 1)^{(n)}$, we identify the series expansion of the confluent hypergeometric function defined as
\begin{equation}
    {}_1F_1(A;B;z) =
    \sum \frac{A^{(n)}}{B^{(n)}} \frac{z^n}{n!},
\end{equation}
so that
\begin{equation}
    \int t^s e^{-k t}\, \mathrm{d}t =
    \frac{t^{s + 1}}{s + 1} {}_1F_1(s + 1; s + 2; -kt).
    \label{eq:integral_primitive}
\end{equation}
Finally, we apply the primitive to our two terms $\delta a^{p+1}$ and $\eta a^{p+2}$ and evaluate at the integration limits $[1, a]$:
\begin{equation}
    I(a) = 
    e^k \left[ \frac{\delta}{p + 1} J(p + 1, a) + \frac{\eta}{p + 2} J(p + 2, a) \right],
\end{equation}
where
\begin{equation}
    J(s, a) = a^s {}_1F_1(s; s + 1; -ka) - {}_1F_1(s; s + 1; -k).
\end{equation}

\section{\label{sec:evidence_correction} \texorpdfstring{Unbiased estimation \\ of the Bayesian evidence}{Unbiased estimation of the Bayesian evidence}}

The uniform priors defined in Table~\ref{tab:priors} span a parameter space that inevitably encompasses regions violating physical viability. Specifically, the IDE models must satisfy the background stability conditions, given by~\eqref{eq:early_time_constraint}, and $H(z)^2 \geq 0$. The latter cannot be clearly mapped into a well-defined prior distribution that can then be passed to a Nested Sampling algorithm. For this reason, we enforce these constraints during sampling by assigning a null likelihood ($\mathcal{L} = 0$) to any parameter combination that violates physical viability, i.e., we follow a rejection sampling approach as in~\cite{Ong:2025ver}.

This procedure effectively truncates the original prior volume to a smaller, physically valid fractional volume $1 - f_\mathrm{forb}$, where $f_\mathrm{forb}$ is the forbidden fraction of the parameter space. Because \texttt{Nautilus} explores and normalizes its Bayesian evidence estimate $\mathcal{Z}_\texttt{Nautilus}$ over the full user-defined prior space, we must correct this raw value to obtain the true physical evidence $\mathcal{Z}$, properly normalized over the valid region:
\begin{equation}
    \ln \mathcal{Z} =
    \ln \mathcal{Z}_\texttt{Nautilus} - \ln (1 - f_\mathrm{forb}).
\end{equation}

In practice, we evaluate the valid volume fraction $(1 - f_\mathrm{forb})$ via Monte Carlo integration, drawing samples from the uniform prior bounds and computing the fraction that satisfies all physical constraints. To ensure this volume correction does not degrade the statistical precision of our model comparison, we draw a sufficiently large number of samples such that the associated fractional error remains subdominant to the inherent target evidence uncertainty of \texttt{Nautilus} ($\Delta \ln \mathcal{Z} \approx 0.01$)~\cite{Lange:2023ydq}. All Bayes factors represented in Figure~\ref{fig:evidence} incorporate this exact volume correction, and in most cases account for a correction of $\mathcal{O}(1)$ in $\ln \mathcal{Z}$. In any case, from multiple runs we expect that the error in the reported values are of order $\mathcal{O}(0.1)$.

\bibliographystyle{apsrev4-2-titles}
\bibliography{Bibliography}

\end{document}